\documentclass{kapproc} 

\usepackage{t1enc}

\usepackage{procps} 

\usepackage[dvips]{graphicx}
\upperandlowercase

\kluwerbib 

\begin{document}



\articletitle[Properties of Ly-{$\alpha$} and Gamma Ray Burst selected 
starbursts at high redshifts] 
{Properties of Ly-{$\alpha$} and Gamma Ray Burst 
selected starbursts at high redshifts}
\author{
Johan P. U. Fynbo\altaffilmark{1}, 
Brian Krog\altaffilmark{2,3},
Kim Nilsson\altaffilmark{1}, 
Gunnlaugur Bj{\"o}rnsson\altaffilmark{4},
Jens Hjorth\altaffilmark{1},
P{\'a}ll Jakobsson\altaffilmark{1,4},
C{\'e}dric Ledoux\altaffilmark{5},
Palle M\o ller\altaffilmark{6},
Bjarne Thomsen\altaffilmark{3} 
} 

\affil{
\altaffilmark{1}Niels Bohr Institute, University of Copenhagen, Denmark\\
\altaffilmark{2}Nordic Optical Telescope, Santa Cruz de La Palma, Spain\\
\altaffilmark{3}Department of Physics and Astronomy, University of \AA rhus,
Denmark\\
\altaffilmark{4}Science Institute, University of Iceland, Iceland.\\
\altaffilmark{5}European Southern Observatory, Casilla 19001, Santiago 19, Chile\\
\altaffilmark{6}European Southern Observatory, Garching bei M\"unchen, Germany
}  

\begin{abstract}
Selection of starbursts through either deep narrow band imaging of
redshifted Ly-$\alpha$ emitters, or localisation of host galaxies of gamma-ray
bursts both give access to starburst galaxies that are significantly fainter
than what is currently available from selection techniques based on the
continuum (such as Lyman-break selection). We here present results from a
survey for Ly-$\alpha$ emitters at $z=3$, conducted at the European Southern
Observatory's Very Large Telescope. Furthermore, we briefly describe the
properties of host galaxies of gamma-ray bursts at $z>2$. The majority of
both Ly-$\alpha$ and gamma-ray burst selected starbursts are fainter than the
flux limit of the Lyman-break galaxy sample, suggesting that a significant
fraction of the integrated star formation at $z\approx3$ is located in galaxies
at the faint end of the luminosity function.  \end{abstract}


\section{Introduction}
As illustrated in the title of this conference, {\it From 30-Doradus to
Lyman-break galaxies}, the term {\it Lyman-break galaxy} (LBG) is almost 
synonymous with {\it high redshift starburst galaxy}. However, 
as has been stressed by many authors, including the lead researchers 
behind the Lyman-break technique, current samples of LBGs
consist of starbursts that are extremely luminous
in the UV and do not give a complete census of all starbursts at high 
redshifts. The current magnitude limit in the ground based surveys for
LBGs is R(AB)$=25.5$ (Steidel et al.\ 2003). The luminosity function of 
LBGs has been extended to R(AB)$\approx27$ based on data from the Hubble Deep 
Fields (e.g., Adelberger \& Steidel 2000). The faint end of this LBG
luminosity function is very steep, with a slope $\alpha = -1.6$,
implying that more than 70\% of the light is emitted by galaxies fainter
than R(AB)$=25.5$.  
Furthermore, as discussed at this conference, the most vigorous starbursts at
high redshifts, e.g. as observed with SCUBA or with Spitzer, are often obscured
in the rest-frame UV (e.g., Chapman et al.\ 2004) and often hence do not
fulfill the selection criteria for LBGs.

How is it then possible to locate and examine starbursts at high redshifts that
are missed by the Lyman-break technique? One other method, not mentioned at
this conference, is absorption selection of galaxies. The few galaxy
counterparts that so far have been identified for Damped Ly-$\alpha$ Absorbers
(DLAs),
found in QSO spectra, appear to be starburst galaxies with significantly
smaller luminosities than LBGs (e.g., M\o ller et al.\ 2004; 
Weatherley et al.\ 2005). However, the total cross section of DLAs at
$z\approx3$ is much larger than what can be accounted for by LBGs. This
implies that most of the neutral gas available for star formation at these
redshifts is located in galaxies fainter than the LBG flux limit (Fynbo et al.
1999).  Other papers in these proceedings discuss the dust emission selected
galaxies. Here we will discuss {\it i)} selection of Ly-$\alpha$ emitting
starbursts through deep narrow band imaging, and {\it ii)}
localisation of gamma-ray burst (GRB) host galaxies.

\section{Ly-$\alpha$ selection of high redshift starbursts}
The idea to use Ly-$\alpha$ to search for primordial galaxies dates back to
the 1960s (Partridge \& Peebles 1967). The first detection of redshifted
Ly-$\alpha$ emission from galaxies not powered by active galactic nuclei
(AGN) were serendipitous discoveries resulting from searches for galaxy
counterparts of QSO absorbers such as DLAs and Lyman-limit systems (Lowenthal
et al.\ 1991; M\o ller \& Warren 1993; Francis et al.\ 1996; Fynbo et al.\ 2001;
Francis et al.\ 2004). Other Ly-$\alpha$ emitters were serendipitously
discovered during searches for intra-cluster planetary nebulae (Kudritzki et
al. 2000).  This curiosity reflects the fact that for many years it was
considered impossible to locate galaxies by their Ly-$\alpha$ emission as
the probability for Ly-$\alpha$ photons to be absorbed by dust, due to resonant
scattering, is much larger than for continuum photons.  The first dedicated
search for Ly-$\alpha$ emitters with 8--10 m class telescopes was conducted at
the Keck telescope (Hu et al.\ 1998).

\subsection{The ``Building the Bridge'' Survey} 
\begin{figure}
\begin{center}
\includegraphics[width=0.85\textwidth]{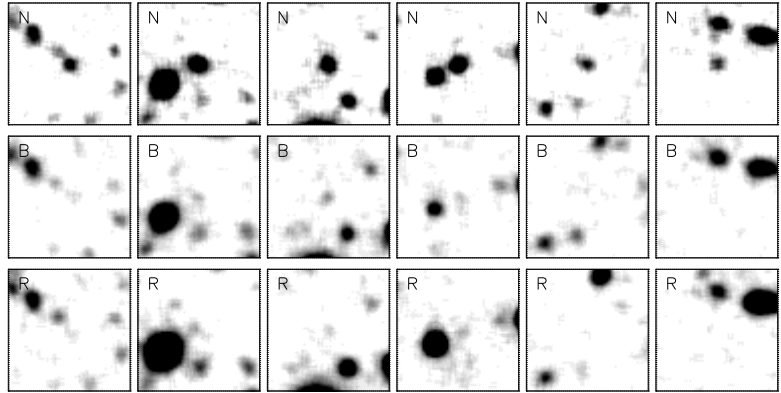}
\vskip -0.2cm
\caption{ Six examples of spectroscopically confirmed $z=2.85$ Ly-$\alpha$
emitters from the ``Building the Bridge Survey'' (see Fynbo et al.\ 2003a for
more information including spectra of the sources). The size of each image is
$12\times 12$ arcsec$^2$. The upper row shows narrow band images of the
galaxies, middle row the B-band images and the bottom row the R-band images.
About $\sim 20$\% of the LEGOs remain undetected in the broad bands despite
$5\sigma$ detection limits of ${\rm B}({\rm AB})=27.0$ and ${\rm R}({\rm
AB})=26.4$.
} 
\label{examples}
\end{center}
\end{figure}

In 2000, some of the authors
started the program ``Building the Bridge between
Damped Ly-$\alpha$ Absorbers and Lyman-Break Galaxies: Ly-$\alpha$
Selection of Galaxies'' at the European Southern Observatory's Very Large
Telescope (VLT). This project is an attempt to use Ly-$\alpha$ selection to
bridge the gap between absorption- and emission-line selected galaxies by
characterisation of $z\approx 3$ galaxies, possibly corresponding to the
abundant population of faint (${\rm R}>25.5$) galaxies associated to DLAs
(Fynbo et al.\ 1999; Haehnelt et al.\ 2000).  The survey consists of very deep
narrow band observations of three fields at $z=2.85$, $z=3.15$, and $z=3.20$.
In each of these fields, we have detected and spectroscopically confirmed
$\sim$ 20 Ly-$\alpha$ emitters, or LEGOs (acronym for {\it Ly-$\alpha$ Emitting
Galaxy-building Object}). In Fig.~\ref{examples} we show six examples of LEGOs
from the $z=2.85$ field.  Of the total sample, 85\% are fainter than the flux
limit for LBGs. Furthermore, as only $\approx$25\% of the LBGs have Ly-$\alpha$
emission lines with equivalent widths large enough to fulfill our selection
criterion for LEGOs (Shapley et al.\  2003), it is clear that LBGs and LEGOs are
almost disjunct classes of high redshift starbursts.

\subsection{LEGOs in the GOODS Field South}
Given that most LEGOs are extremely faint, it is very difficult to establish
any property beyond the Ly-$\alpha$ flux for individual galaxies, even with
8--10 m class telescopes. For this reason, we decided to observe a field with
existing, very deep broad band observations covering most of the
electromagnetic spectrum, namely the GOODS Field South (Giavalisco et al.
2004). In 2002 we obtained observations of a section of the GOODS
Field South. Due to bad weather, we did not reach the same flux limit as for
the ``Building the Bridge'' fields, but still we detected nearly 20 candidate
$z=3.20$ LEGOs in the field.

\begin{figure}
\begin{center}
\includegraphics[width=0.90\textwidth]{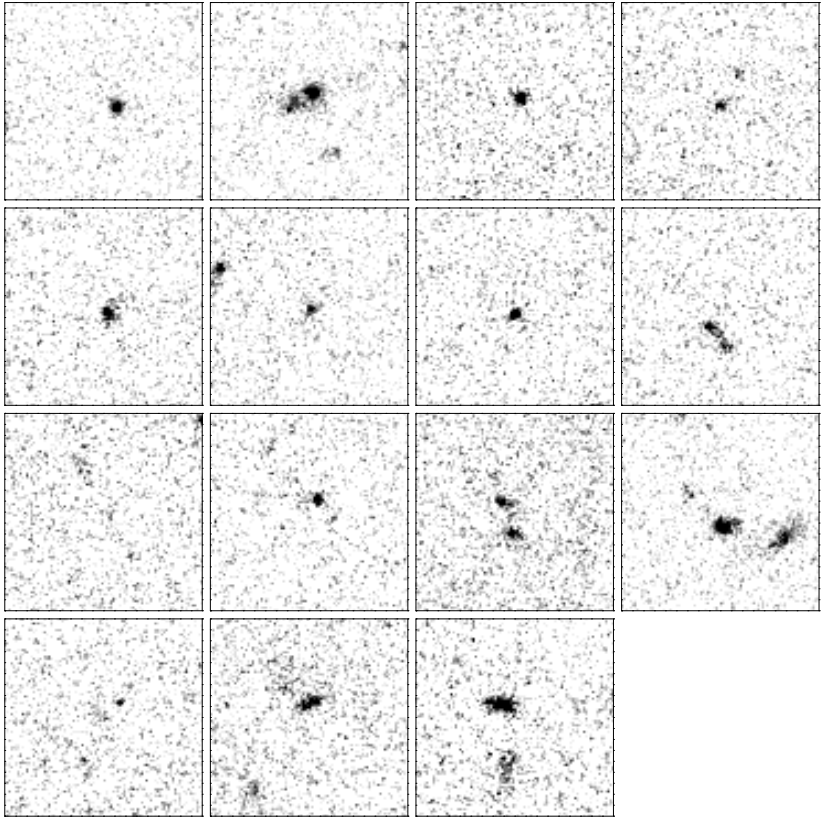}
\vskip -0.2cm
\caption{
HST/ACS V-band images of size 3$\times3$ arcsec$^2$ around the 
positions of 15 of the candidate $z=3.20$ LEGOs in part of the GOODS Field South
.  The V(AB) magnitudes range from $\sim$26 to $>$29.
} 
\label{goods}
\end{center}
\end{figure}

The analysis of the broad band properties of these candidates constitute the
thesis work of two of the authors (B. Krog and K. Nilsson). Here we report a
few preliminary results. So far the objects have been studied in X-rays
(Chandra X-ray Observatory, 1 Ms exposure), near-IR (VLT/ISAAC), and the
optical broad bands (HST/ACS). The HST images (Fig.~\ref{goods}) confirm that
the LEGOs 
have extremely faint continua in the range V(AB)$=26$--29.
Furthermore, {\it i)} these galaxies are bluer than most LBGs, with spectra that
rise toward the blue ($F_{\nu} \propto \nu^{\beta}, \beta < 0$) implying
younger ages and/or lower dust content than what is typical for LBGs, and
{\it ii)} they have extremely compact morphologies and sometimes consist
of several knots similar to the LBGs (e.g., Lowenthal et al.\ 1997). Only 
upper limits were found in the X-ray and near-IR images so we can exclude that 
the galaxies harbour AGN and significant older populations of stars. 
We are currently working on deriving stronger constraints from stacking 
of the individual sources.

\section{GRB selection of starbursts} 
GRBs are short, extremely energetic bursts of $\gamma$-rays
associated with energetic core-collapse supernovae (Hjorth et al.\ 2003). 
If the GRB rate is directly proportional to the (massive) star
formation rate, then the properties of GRB hosts should reflect the diversity
of all star-forming galaxies in terms of luminosity, environment, internal
extinction and star formation rate. GRB hosts therefore constitute a central
clue for our understanding of galaxy formation and evolution (e.g., 
Ramirez-Ruiz et al.\ 2002; Tanvir et al.\ 2004; see also the papers by Tanvir 
et al.\ and Trentham et al.\ in these proceedings).

The prompt burst of $\gamma$-rays is followed by a so-called afterglow,
emitting over a very wide spectral range from radio through the optical/near-IR
to X-rays (see van Paradijs et al.\ 2000 for a review). The afterglow can be
extremely bright, allowing a precise localisation on the sky.
More importantly, spectroscopy of the afterglow can
give information about the redshift and the physical conditions within the
host, and in intervening absorption systems along the line of sight (see,
e.g., Vreeswijk et al.\ 2004 and Jakobsson et al.\ 2004a for examples). When the
GRB has faded, the host galaxy itself can be observed. The measured redshifts
for GRBs cover a very broad range from $z=0.0085$ to $z=4.50$ with a median
around $z=1.1$ (e.g., Jakobsson et al.\ 2004a).  

\begin{figure}[t]
\begin{center}
\includegraphics[width=0.351\textwidth]{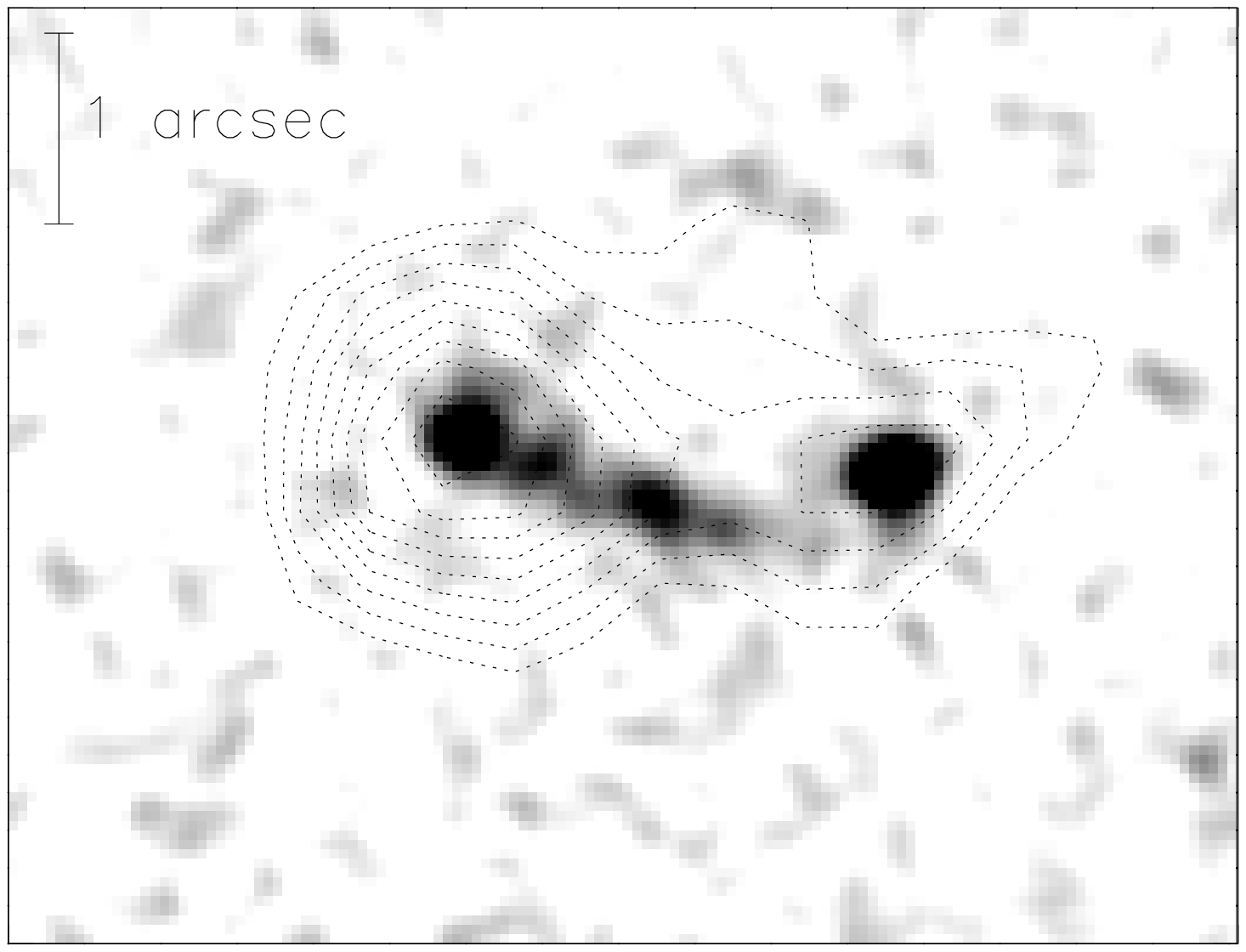}
\includegraphics[width=0.267\textwidth]{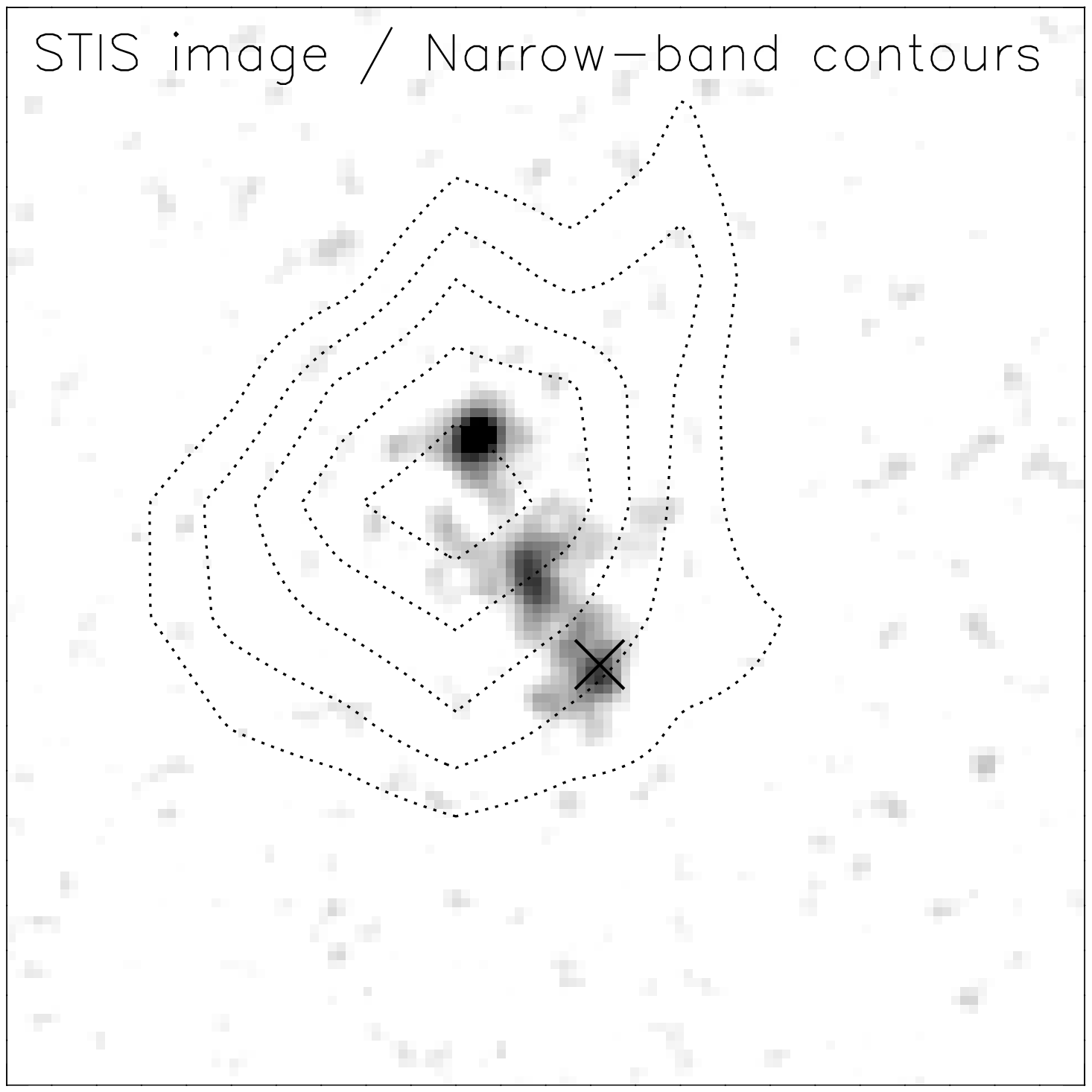}
\includegraphics[width=0.267\textwidth]{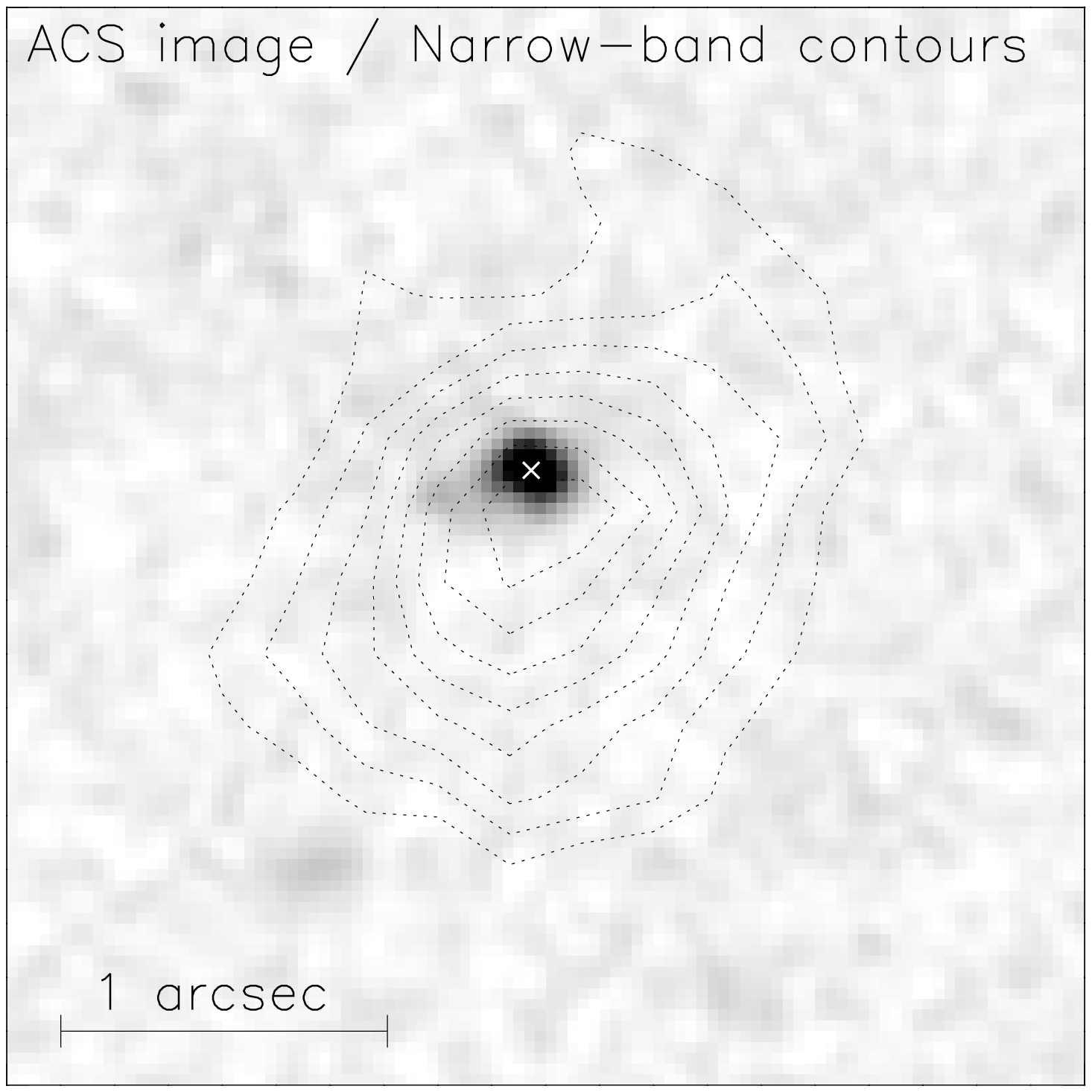}
\vskip -0.2cm
\caption{
HST images of the host galaxies of GRBs 000926 ($z=2.04$, R(AB)$=24.0$), 011211
($z=2.14$, R(AB)$=25.0$), and 021004 ($z=2.33$, R(AB)$=24.4$). The contours
show the morphology of Ly-$\alpha$ emission, as measured by ground-based narrow
band imaging (Fynbo et al.\ 2002; Fynbo et al.\ 2003b; Jakobsson et al.\, in
prep). GRB~000926 occurred near the centre of the right-most knot. The
locations of GRB~011211 and GRB~021004 are marked with crosses. The field sizes
for the GRB~011211 and GRB~021004 images are the same.
}
\label{hosts}
\end{center}
\end{figure}

How do GRB hosts compare to LBGs and LEGOs?  So far, the redshift has been
measured for 10 $z>2$ GRBs. HST-images of the three brightest of these are
shown in Fig.~\ref{hosts}. For the remaining seven, the host is either
undetected to faint magnitude limits or have a magnitude fainter than
R(AB)$=26$, so the majority of these GRB hosts are fainter than LBGs.
Remarkably, all current evidence is consistent with the conjecture that GRB
hosts are Ly-$\alpha$ emitters (Fynbo et al.\ 2003b). As shown in
Fig.~\ref{hosts}, the three brightest high redshift GRB hosts are Ly-$\alpha$
emitters, and for the remaining seven, Ly-$\alpha$ emission has either been
detected (2 hosts) or not yet searched for to sufficient depth to allow
detection of even a large equivalent width emission line (5 hosts).  Taken at
face value, this suggest that faint, LEGO-like galaxies in total account for
the majority of the star formation at these redshifts. However, there could be
other explanations for why GRBs have, so far, mainly been localised in such
galaxies (Fynbo et al.\ 2003b). In particular, some of the so-called dark bursts
could be located in more massive and dust-obscured galaxies (e.g., Tanvir et al
2004; Jakobsson et al.\ 2004b and references therein). The recently launched
Swift satellite (Gehrels et al.\ 2004) offers a unique chance to resolve this
issue.

\section{Summary} In conclusion, surveys for LEGOs and for GRB host
galaxies reveal that a major fraction of the starburst activity
at $z>2$ may be located in galaxies fainter than the flux limit of 
the LBG survey. We have here shown that Ly-$\alpha$ emission and 
GRB selection are two viable methods to probe this population
of faint starbursts at high redshifts.

\begin{acknowledgments} 
We acknowledge benefits from collaboration within the EU FP5 Research
Training Network ``Gamma-Ray Bursts: An Enigma and a Tool''.
This work is supported by the Danish Natural Science Research
Council (SNF).
\end{acknowledgments}

\end{document}